\documentclass[aps]{revtex4}%
\usepackage{amsfonts}
\usepackage{amsmath}
\usepackage{amssymb}
\usepackage{graphicx}%
\begin{document}
\preprint{ }
\title[Incommensurate phase in thin films]{Phenomenological approach to a transition into the \\incommensurate phase for thin films}
\author{E.V. Charnaja}
\affiliation{Institute of Physics, St. Petersburg State University,\\
St. Petersburg, 198504, Russia}
\author{S.A. Ktitorov}
\affiliation{A.F. Ioffe Physico-Technical Institute\\ of the Russian 
Academy of Sciences, Polytechnicheskaja str. 26,\\ St. Petersburg 194021, Russia}
\author{O.S. Pogorelova}
\affiliation{I.I. Mechnikov Medical University,\\ 
St. Petersburg, 195067, Russia}

\begin{abstract}
A phenomenological theory of the incommensurate phase transition in thin films
is presented for the case of the single-component order parameter. The order
parameter distribution over the film is obtained under the assumption of
slowly-varying amplitude. The size effect on the temperature of the phase
transition into the incommensurate phase has been considered.

\end{abstract}
\maketitle



\section{Introduction}

Numerous recent experimental studies of thin films and small particles
manufactured on the base of ferroelectric materials, such as BaTiO$_{3}$,
PbTiO$_{3}$ and some others (see, for example, [1-8] and references therein)
have shown that temperatures of phase transitions into a polarized state
change noticeably (mostly, being reduced) with particle size or film thickness
decrease, the phase transitions smeare out and the ferroelectric state is
normally destroyed at some particular linear size. Spontaneous polarization,
susceptibility, coercive and switching fields are also affected. Most of these
features could be explained within theoretical models developed recently for
ferroelectrics (see [9-15]). In particular, a phenomenological approach to the
ferroelectric phase transitions in thin films and small particles was
described in [9-11], while the Ising model was used in [12, 13]. The
antiferroelectric phase transition in small particles and thin films was
considered later in [17] on the base of the Landau theory. In all papers cited
above, the polar states in relevant bulk materials were nonmodulated and the
spatial variation of the spontaneous polarization and other parameters was
induced by surface only (disregarding fluctuations). It is of interest to
extend the phenomenological description of phase transitions in
low-dimensional systems to transitions into modulated states; in particular,
into the incommensurate phase. Note that some experimental studies of
incommensurate thin films were made recently in [18]. The main difficulty of
the phenomenological consideration of the size effect on the incommensurate
phase transition is due to the fact that the incommensurate phase in the bulk
is described by a nonlinear fourth-order differential equation for the order
parameter contrary to the case of ferroelectrics or antiferroelectrics [19].
Thus, the boundary conditions for the order parameter, which were first
suggested by de Gennes [20] and then were generally used in phenomenological
approaches to the size effect, do not suffice. However, as it will be shown
below, some assumptions allow us to reduce the problem to a simpler
differential equation and to analyze the incommensurate phase transition
quantitatively. The boundary conditions for this approximate equation are
deduced from the conditions for the fourth-order one.

In the present paper we suggest for the first time a phenomenological model of
the transition into the incommensurate phase in thin films.\bigskip

\section{Theoretical model}

Let us consider a thin film of thickness $L$. We assume that the $x$ axis of
the coordinate system is perpendicular to the film surfaces and the origin
$x=0$ coincides with the film middle. Notice that due to the geometry chosen,
the order parameter depends only on $x$. The order parameter $\eta$ can be the
spontaneous polarization if the sample undergoes a transition into the
ferroelectric phase at lower temperatures, but the further treatment does not
depend significantly on its physical nature and may be applied to
incommensurate phases of any physical nature. The functional detrmining the
equilibrium equation and the boundary condition can be written in the form%
\begin{align}
\Phi &  =\int_{-L/2}^{L/2}dx\left[  \frac{1}{2}A_{0}\left(  T-T_{0}\right)
\eta^{2}+\frac{1}{4}B\eta^{4}+\frac{1}{2}C\left(  \frac{d\eta}{dX}\right)
^{2}+\frac{1}{2}D\left(  \frac{d^{2}\eta}{dX^{2}}\right)  ^{2}\right]
+\nonumber\\
&  \psi(\eta(L/2)-\phi(\eta(-L/2)), \label{potent}%
\end{align}
where $A_{0},B,C,D$ are phenomenological parameters, $A_{0}>0,$ $B>0,$ $C<0,$
$D>0,$ $T_{0}$ is the characteristic temperature; the functions $\psi
(\eta(L/2))$ and $\phi(\eta(-L/2))$ determine the boundary conditions [22].
Having in mind that the symmetric solutions give usually the lowest magnitude
of the thermodynamic potential in the case of symmetric samples, we consider
below only even functions of $x$ for the order parameter distribution:
$\eta(-x)=\eta(x)$ and only the region of $x\geq0$ will be considered
explicitly. Varying the functional (\ref{potent}), we obtain the Euler
equation [21]:%

\begin{equation}
\frac{d^{4}\eta}{dx^{4}}-\frac{C}{D}\frac{d^{2}\eta}{dx^{2}}+\frac{B}{D}%
\eta^{3}+\frac{A}{D}\eta=0 \label{euler}%
\end{equation}
describing the inhomogeneous equilibrium state and the natural boundary
conditions [21]%

\begin{equation}
\left[  \frac{d\psi(\eta)}{d\eta}+\frac{dF}{d\eta^{\prime}}+\frac{d}{dx}%
\frac{\partial F}{\partial\eta^{\prime\prime}}\right]  _{x=L/2}=0,
\label{bound2}%
\end{equation}

\begin{equation}
\left[  \frac{\partial F}{\partial\eta^{\prime\prime}}\right]  _{L/2}=0,
\label{bound6}%
\end{equation}
where $F(\eta(x),\eta^{\prime},\eta^{\prime\prime})$ is the integrand of the
functional (\ref{potent}), $\eta^{\prime}\equiv\frac{d\eta}{dx},$
$\eta^{\prime\prime}\equiv\frac{d^{2}\eta}{dx^{2}}$. We choose the simplest
relevant form for $\psi$:%

\begin{equation}
\psi(\eta(L/2))=\alpha\frac{\eta^{2}}{2}. \label{bound3}%
\end{equation}
Using the boundary conditions(\ref{bound2}) and (\ref{bound3}), and the
functional (\ref{potent}), we obtain:%

\begin{equation}
\left[  \alpha\eta(x)+C\frac{d\eta(x)}{dx}+D\frac{d^{3}\eta(x)}{dx^{3}%
}\right]  _{L/2}=0, \label{bound4}%
\end{equation}%
\begin{equation}
\left[  \frac{d^{2}\eta(x)}{dx^{2}}\right]  _{L/2}=0. \label{bound5}%
\end{equation}
These boundary conditions can be used for numerical solving of the
fourth-order equation (\ref{euler}). However, we consider an approximate
reduction of this equation to the second-order one, which will be solved exactly.

\section{Slowly-varying amplitude}

It is known that the order parameter in the bulk incommensurate phase
oscillates with a steady amplitude according to the equation [19] $\eta
=u\cos(qx),$ where the oscillation wave vector $q$ is given by the relation%

\begin{equation}
q=\sqrt{-\frac{C}{2D}}. \label{q}%
\end{equation}
When the oscillation period in the bulk \ is implied to be much shorter than
the film thickness, i. e. $\left\vert D/C\right\vert \ll L^{2}$, we can assume
that the restricted geometry does not influences noticeably the period of the
order parameter modulation and yields only alterations in the amplitude $u(x)$
as a function of $x$. Then the $x$ dependence of $\eta$ in the film can be
written as%

\begin{equation}
\eta(x)=u(x)\cos(qx). \label{amplit}%
\end{equation}
Under the assumption of $\left\vert D/C\right\vert \ll L^{2}$\ it is
reasonable to assume in the spirit of the slowly-varying amplitude approach
that variation of $u(x)$ is smooth enough and its derivatives of order higher
than the second one can be omitted. Then the relationship (\ref{amplit}) leads
to the following second-order differential equation for the amplitude $u$:%

\begin{equation}
\frac{d^{2}u}{dx^{2}}=-\frac{3B}{8C}u^{3}+\frac{C^{2}-4AD}{8CD}u.
\label{2order}%
\end{equation}
This equation along with relevant boundary conditions yields a boundary value
problem for the amplitude $u$ within the thin film. The boundary conditions
for $u$ can be deduced from the conditions for the fourth-order equation.

\section{Boundary conditions for the second-order equation}

The equation (\ref{2order}) must be supplemented with boundary conditions that
can be derived from (\ref{bound4}) and (\ref{bound5}) using the slowly-varying
amplitude approach. We substitute the expression (\ref{amplit}) into these
conditions and neglecting higher derivatives of $u(x)$ we obtain a set of equations:%

\begin{equation}
\left[  (C+3Dq^{2})\left(  \frac{u^{\prime}}{u}\right)  +\alpha\right]
\cos\left(  \frac{qL}{2}\right)  =q\left(  C+Dq^{2}\right)  \sin\left(
\frac{qL}{2}\right)  , \label{set1}%
\end{equation}

\begin{equation}
q^{2}\cos\left(  \frac{qL}{2}\right)  =-2q\left(  \frac{u^{\prime}}{u}\right)
\sin\left(  \frac{qL}{2}\right)  . \label{set2}%
\end{equation}
Eliminating the sine and cosine functions from (\ref{set1}) and (\ref{set2})
and introducing the following notation for the logarithmic derivative,
\begin{equation}
\frac{u^{\prime}(L/2)}{u(L/2)}=-1/\delta, \label{bcquadro}%
\end{equation}
we obtain a quadratic equation for the logarithmic derivative:%

\begin{equation}
\left(  C+Dq^{2}\right)  \frac{q^{2}}{2}\delta^{2}-\alpha\delta+\left(
C+3Dq^{2}\right)  =0. \label{quadrat}%
\end{equation}
Its solution reads%
\begin{equation}
\delta=\frac{\alpha\mp\sqrt{\alpha^{2}-2\left(  C+Dq^{2}\right)  \left(
C+3Dq^{2}\right)  q^{2}}}{q^{2}\left(  C+Dq^{2}\right)  }. \label{quadrsol}%
\end{equation}
The relation (\ref{bcquadro}) plays a role of the boundary condition for the
equation (\ref{2order}) for the amplitude. It has a form similar to the
mentioned in the Introduction boundary condition of de Gennes. Real-valuedness
of $\delta$ implies that $\alpha^{2}-2\left(  C+Dq^{2}\right)  \left(
C+3Dq^{2}\right)  q^{2}\geq0$. Taking for the $q$ value the expression
(\ref{q}), we obtain%

\begin{equation}
\delta=-\frac{4D}{C^{2}}\left(  \alpha\pm\sqrt{\alpha^{2}-\frac{C^{3}}{4D}%
}\right)  . \label{delta}%
\end{equation}
\bigskip The discriminant $\alpha^{2}-\frac{C^{3}}{4D}$ is always positive
since $C<0,D>0$; therefore, $\delta$ is always real. The two roots of
(\ref{quadrat}) have opposite signes, but only one of them has a physical
meaning at a given magnitude of $qL$. Indeed, compatibility of equations
(\ref{set1}) and (\ref{set2}) implies that%

\begin{equation}
sign\delta=sign\tan\frac{qL}{2}. \label{sign}%
\end{equation}
This consideration suggests that the magnitude of $q$ is rigidly determined by
the condition (\ref{q}). We have in this case an oscillating dependence of the
amplitude logarithmic derivative at the boundary on the the "sample phase
thickness" $qL$ and, therefore, on the temperature. A behaviour of the system
is ruled really by more complicated equations than our ones obtained within
the slowly-varying amplitude approach. This means that the real system can
violate some of our assumtions in the boundary layer in search of the energy
minimum. One way is to tune $q$ so that the energy is minimum. For thick
samples $qL$ is rather large and only a very small correction of $q$ would
suffice. Another possibility is to change the wave number $q$ only within some
boundary layer. Only numerical analysis of the fourth-order equation can
answer the question: which way is chosen by the system. It is easy to see that
under the assumptions made the problem for the incommensurate phase transition
is reduced to a boundary value problem which is very similar to that for the
ferroelectric thin film. The surface extrapolation length $\delta$ in
(\ref{bcquadro}) describes the difference between the surface and the bulk and
can be positive or negative. It is natural to suppose that for free thin films
$\delta$ should be positive. Size effect in thin films is noticeable when the
extrapolation length value is comparable with the film thickness. Thus, if $L$
is assumed to be much longer than the bulk modulation period $2\pi/q$, the
same is valid for $\delta$: $\delta\gg2\pi/q$. We consider in the next section
the analytic solution for the second-order equation.

\section{Analytic solution}

It is fruitful to analyze the phase portrait of the equation (4), which is
presented in Fig. 1.%

\begin{figure}
[p]
\begin{center}
\includegraphics[
trim=0.000000pt -33.739876pt 11.046865pt 50.359901pt,
height=450.625pt,
width=350pt
]%
{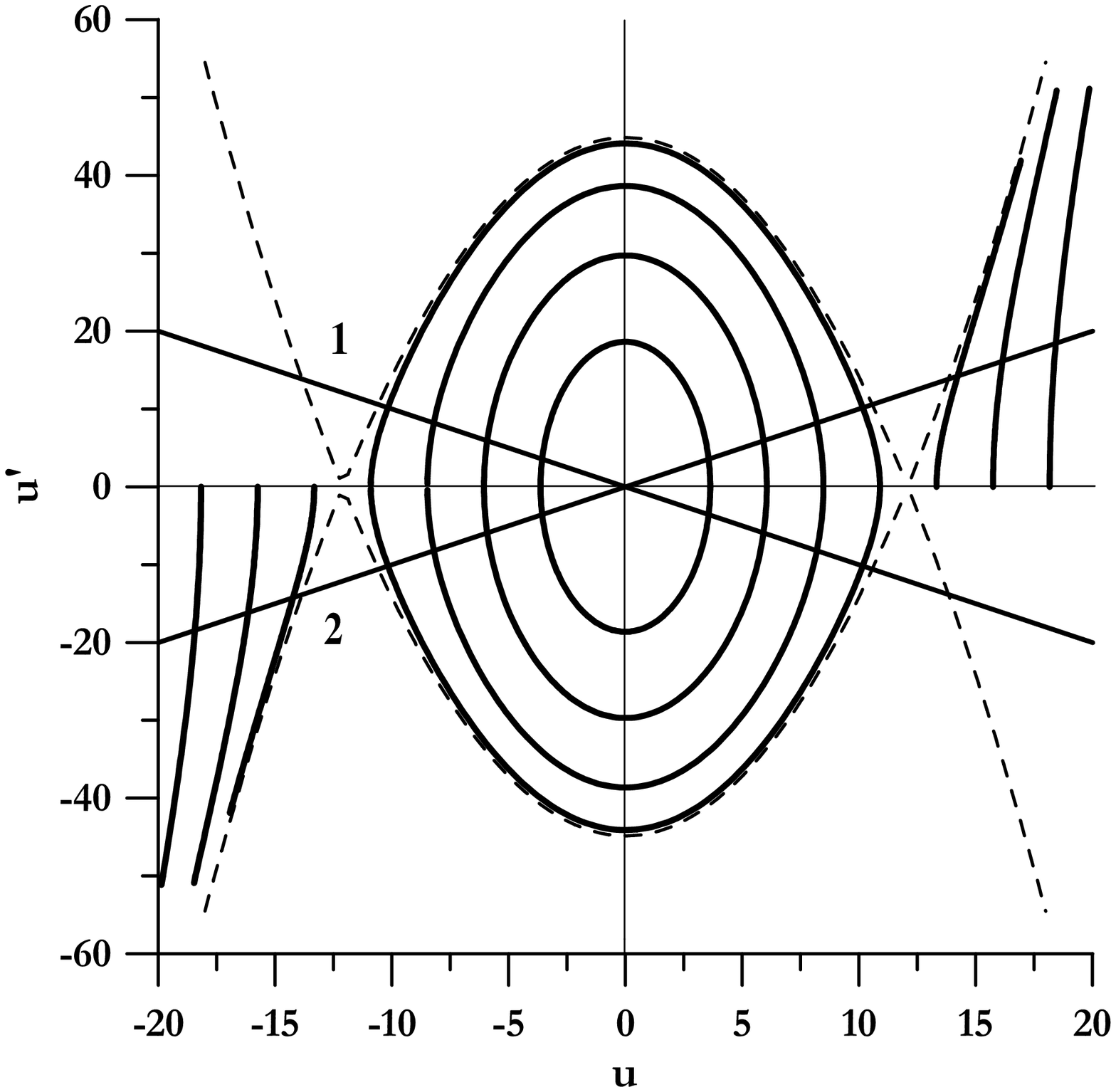}%
\caption{The phase portrait for eq. 4. A=1000, B=100, C=200, D-1, T$_{0}$=1.
The dashed lines are the separatrices. The straight lines correspond to the
boundary condition (7) at $\delta$=1 (1) and $\delta$=-1 (2).}%
\label{fig.1}%
\end{center}
\end{figure}

One can see two regions of solutions separated by separatrices which are shown
by dashed lines in Fig. 1. The enclosed curves, which correspond to periodic
solutions of (4), lie within the separatrices. The boundary condition eq. (7)
look on the phase diagram as a straight line with negative slope for positive
extrapolation lengths and vice versa. It is seen from Fig.1 that for positive
extrapolation lengths all possible solutions lie within the region enclosed by
the separatrices.

The analytic solution of the equation (4) reads%

\begin{equation}
u\left(  x\right)  =u_{0}\operatorname*{sn}\left(  \sqrt{\frac{4DA-C^{2}}%
{8DC}+\frac{3Bu_{0}^{2}}{16C}}x,k\right)  \label{analytsol}%
\end{equation}
where $\operatorname*{sn}(z,k)$ is the Jacoby elliptic sine function [22],
$u_{0}=u(x=0)$ is the value of $u$ in the film middle, $k$ is the elliptic module:%

\begin{equation}
k=\frac{u_{0}}{\sqrt{\frac{4(C^{2}-4DA)}{3DB}-u_{0}^{2}}} \label{module}%
\end{equation}
and $u_{0}$ is related with $\delta$ by the following relationship:%

\begin{equation}
\delta=-\frac{\operatorname{tn}\left(  \sqrt{\frac{4DA-C^{2}}{8DC}%
+\frac{3Bu_{0}^{2}}{16C}}\frac{L}{2},k\right)  }{\sqrt{\frac{4DA-C^{2}}%
{8DC}+\frac{3Bu_{0}^{2}}{16C}}\operatorname{dn}\left(  \sqrt{\frac{4DA-C^{2}%
}{8DC}+\frac{3Bu_{0}^{2}}{16C}}\frac{L}{2},k\right)  } \label{delta2}%
\end{equation}
($\operatorname{tn}$ and $\operatorname{dn}$ are the Jacoby elliptic functions [22]).

\bigskip%
\begin{figure}
[p]
\begin{center}
\includegraphics[
trim=0.000000pt 0.000000pt -33.536255pt -30.972723pt,
height=450.875pt,
width=350.125pt
]%
{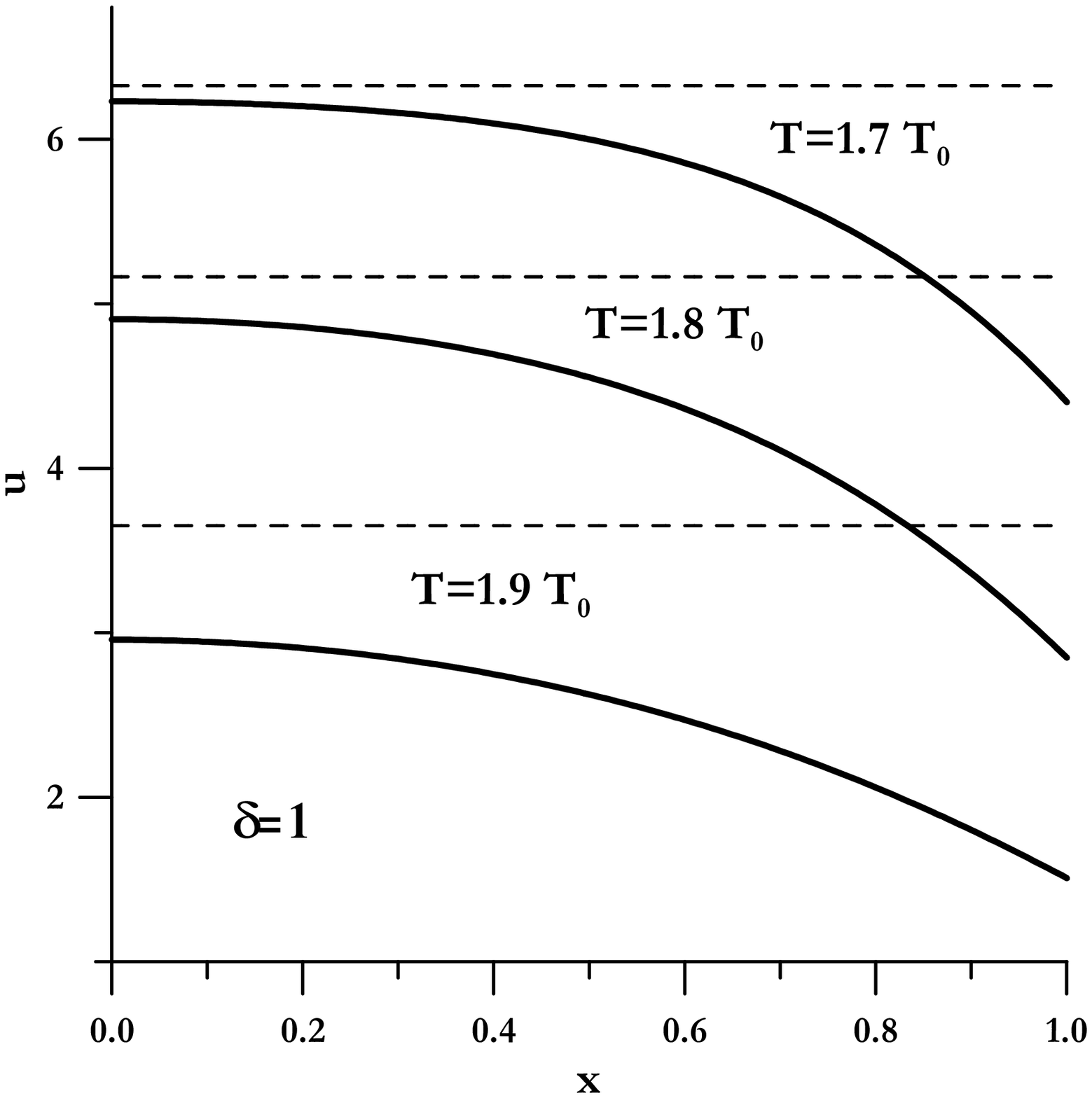}%
\caption{The order parameter amplitude $u$ versus $x$ ($0\leq x\leq L/2$) for
positive $\delta$ at various temperatures. $L=2$; other parameters are given
in the caption to Fig. 1. The dashed straight lines show the amplitude in the
bulk at the same temperatures.}%
\end{center}
\end{figure}

The obtained solutions allow us to find a distribution of the order parameter
amplitude over the film for some particular sets of phenomenological
parameters, film thickness, and temperature. They also allow us to find a
dependence of the phase transition temperature $T_{i}$ on the film thickness
for a particular set of the phenomenological parameters, $T_{i}$ stands for
the temperature where $u_{0}$ (and therefore $u(x)$) vanishes.

The order parameter amplitude $u$ versus $x$ is depicted in Figs. 2 and 3 for
some particular positive as well as negative values of $\delta$ for several
magnitudes of the temperature. The phenomenological parameters used are listed
in the figure caption. The general behaviour of $u(x)$ for positive $\delta$
is similar to the behaviour of the spontaneous polarization in ferroelectric
thin films, that is $u$ decreases near the film surfaces and its value in the
film middle is smaller than for the relevant bulk (Fig. 2).

For negative extrapolation lengths there can be two types of solutions for
$u(x)$ in agreement with the phase diagram (Fig. 1). The behaviour of the
solution shown in Fig. 3 is similar to that discussed for ferroelectric thin
films: $u$ increases near the film surfaces and its value in the film middle
is larger than for the relevant bulk.%

\begin{figure}
[p]
\begin{center}
\includegraphics[
height=450.9375pt,
width=350.125pt
]%
{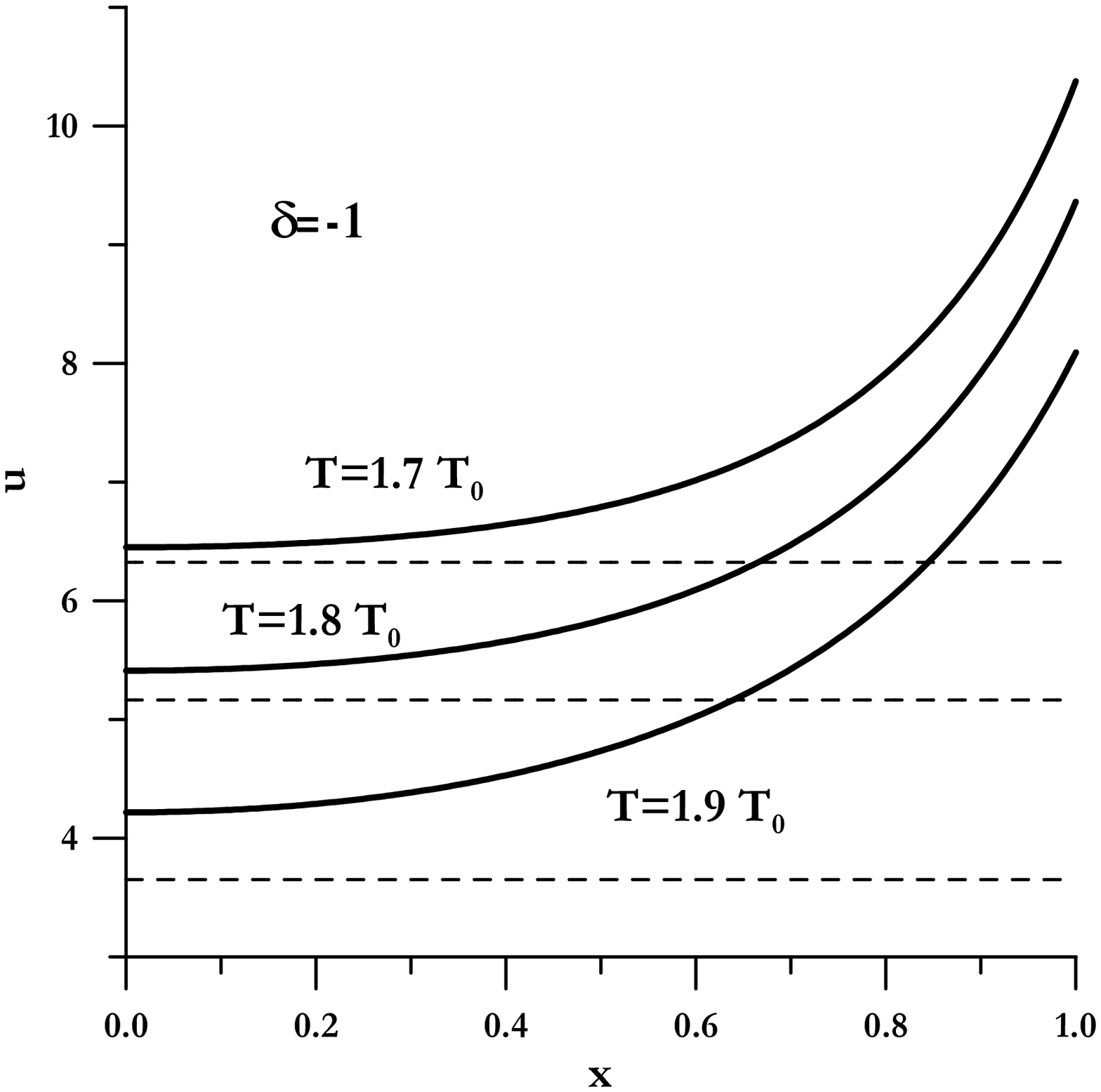}%
\caption{The order parameter amplitude $u$ versus $x$ ($0\leq x\leq L/2$) for
negative $\delta$ at various temperatures. $L=2$; other parameters are given
in the caption to Fig. 1. The dashed straight lines show the amplitude in the
bulk at the same temperatures.}%
\label{fig. 3}%
\end{center}
\end{figure}

The amplitude $u$ within the film for the second solution is smaller than in
the bulk, while its absolute value increases near the surfaces, and the
amplitude changes its sign at some $x$\ that is equivalent to a change of the
incommensurate modulation phase. However, calculation of the equilibrium
thermodynamic potential of the film for two solutions at the negative $\delta$
shows that the first kind of solutions (with $u$ is higher than in the bulk)
is energetically more favourable.%

\begin{figure}
[p]
\begin{center}
\includegraphics[
height=450pt,
width=359.4375pt
]%
{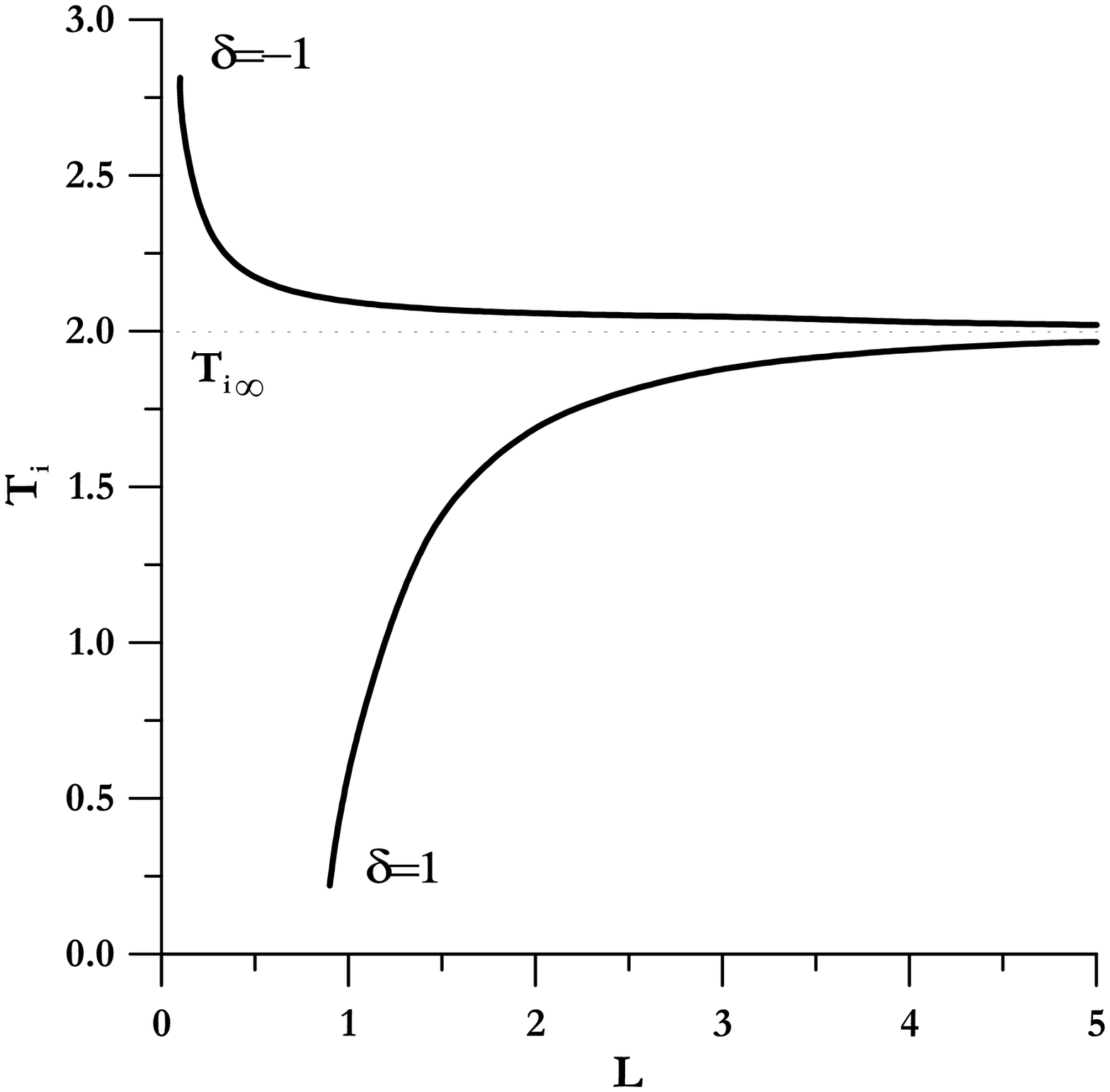}%
\caption{Size dependence of the transition temperature.}%
\label{fig. 4}%
\end{center}
\end{figure}

The size dependence of the incommensurate phase transition temperature for
positive and negative extrapolation lengths is shown in Fig. 4. One can see
that $T_{i}$ decreases with decreasing film thickness at positive $\delta$ and
vanishes at some critical thickness. Similar dependence was obtained for
ferroelectric and antiferroelectric thin films, when $\delta$\ was implied to
be positive [9,17]. The size dependence of $T_{i}$\ for negative $\delta$ is opposite.

\section{Conclusions}

In conclusion, the phenomenological theory of the incommensurate phase
transition in thin films is presented. It was shown that under the assumption
of slowly-varying amplitude of the order parameter the model can be reduced to
a boundary value problem for the second-order differential equation with
conventional homogeneous boundary conditions that can be deduced from ones for
the fourth-order equation. The amplitude of the order parameter was found to
decrease near the film surface for positive extrapolation lengths and to
increase for negative extrapolation lengths. The temperature $T_{i}$ of the
transition was shown to decrease with decreasing the film thickness for
positive extrapolation lengths, the phase transition cannot occur in films
with thickness less than some critical value.

\end{document}